\documentclass[twoside,12pt]{article}
\usepackage{epsfig,bm}

\newcommand{\be}{\begin{equation}}
\newcommand{\ee}{\end{equation}}
\newcommand{\bea}{\begin{eqnarray}}
\newcommand{\eea}{\end{eqnarray}}

\begin{document}

\title{\vspace{1cm} New physics in $B_s^0 \to J/\psi\phi$  decays?}

\author{B.~El-Bennich$^{1,2}$, J.~P.~B.~C.~de Melo$^1$, O.~Leitner$^3$ , B.~Loiseau$^3$ and J.-P.~Dedonder$^3$ \\ \\
$^1$ Laborat\'orio de F\'isica Te\'orica e Computa\c{c}\~ao Cient\'ifica  \\Universidade Cruzeiro do Sul, 01506-000 S\~ao Paulo, SP, Brazil \\
$^2$ Instituto de F\'isica Te\'orica, Universidade Estadual Paulista, \\  01140-070 S\~ao Paulo, SP, Brazil.  \\
$^3$ Laboratoire de Physique Nucl\'eaire et de Hautes \'Energies, \\
          Universit\'e Pierre et Marie Curie et Universit\'e Paris-Diderot, \\ IN2P3 \& CNRS,
          4 place Jussieu, 75252 Paris, France} 

\maketitle

\begin{abstract}
After a brief review of $B_s^0 - \bar B_s^0$ oscillations, we discuss the weak decays $B_s^0 \to J/\psi\phi$ and $B_s^0 \to J/\psi\, f_0(980)$
and the ratio $\mathcal{R}_{f_0/\phi}$ of their decay rates in the light of recent measurements by the LHCb, D$\emptyset$ and CDF Collaborations. 
We point out that the experimental values for $\mathcal{R}_{f_0/\phi}$ impose tight limits on new physics contributions to both decay channels.
\end{abstract}

\section{$\bm{B_s^0 - \bar B_s^0 }$ oscillations into $\bm{J/\psi\,\phi}$ }

The study of $CP$ violation in $B_s$ mesons is still in its early stages with many contemporary experiments and analyses focussing on the decay  $B_s^0 \to J/\psi\phi$. 
The interest in this particular channel owes to the observation that the final state $J/\psi\phi$ is reached by interference of a decay without mixing, $B_s^0 \to  J/\psi\phi$,
and with mixing, $B_s^0 \to \bar B_s^0 \to J/\psi\phi$. In principle, this allows for the observation of the $CP$ violating phase, $-2\beta_s$, where $\beta_s$ is the Standard 
Model angle in the unitarity triangle for the $B_s^0$ system. Practically, it is predicted to be small, $-2\beta_s = \phi_s = -0.038\pm 0.002$~\cite{Lenz:2006hd}, about 
20 times smaller in magnitude than the corresponding phase in $B^0_d$ mixing. 

The mixing occurs in the Standard Model due the nonequivalence of mass and flavor eigenstates and gives rise to particle-antiparticle oscillations. They are described 
by the Cabibbo-Kobayashi-Maskawa (CKM) quark-mixing matrix nowadays established as the leading paradigm for $CP$ violation. The time evolution of $B_s^0$ 
oscillation  follows from the perturbative solution of the time-dependent  Schr\"odinger equation (see, e.g., Ref.~\cite{Anikeev:2001rk} for a detailed review of the formalism),
written as,
\begin{equation}
i \frac{d}{dt} \left(
\begin{array}{c} | B_s^0(t) \rangle \\ | \bar{B}_s^0 (t) \rangle \end{array} \right)  = 
\left( M^s - \frac{i}{2} \Gamma^s \right) \left( \begin{array}{c} | B_s^0(t) \rangle \\ | \bar{B}_s^0 (t) \rangle \end{array} \right) .
\label{schroedinger}
\end{equation}
The complex $2\times 2$ mass and decay rate matrices, $M^s$ and $\Gamma^s$, are hermitian and diagonalization of $M^s - \frac{i}{2} \Gamma^s$ 
yields the mass eigenstates
\begin{eqnarray}
| B_{sL}^0 \rangle & = & p \; | B_s^0 \rangle + q \; | \bar{B}_s^0 \rangle \ ,  \\
| B_{sH}^0 \rangle & = & p \; | B_s^0 \rangle - q \; | \bar{B}_s^0 \rangle \ ,  
\end{eqnarray}
where the masses, $ M^s_L$ and $M^s_H$, and decay rates, $\Gamma_L^s$ and $\Gamma_H^s$, are distinct. The  complex numbers, $p$ and $q$, 
are related to the matrix elements of $M^s$ and $\Gamma^s$ by,
\begin{equation}
  \frac{q}{p}  =  - \frac{2( M_{12}^{s*} -\frac{i}{2}\, \Gamma_{12}^{s*})}{\Delta M^s - \frac{i}{2}\, \Delta \Gamma^s} \ \simeq \
  - \frac{M_{12}^{s*} }{ |M_{12}^s|} \left [ 1 - \frac{1}{2}\, \mathrm{Im}   \left ( \frac{\Gamma_{12}^s}{ M_{12}^s }\right )  \right ] \ , 
 \label{qp}
\end{equation}
and satisfy  $|p|^{2} + |q|^2  =  1$.  The $B_s^0-\bar B_s^0$ oscillations in Eq.~(\ref{schroedinger}) involve the physical quantities $|M^s_{12}|$, 
$|\Gamma^s_{12}|$ and the $CP$ violating phase $\phi_s =\mathrm{arg}( -M_{12}^s/\Gamma_{12}^s)$. The off-diagonal matrix element $\Gamma^s_{12}$ 
is important as it represents the partial width of $B_s^0$ and $\bar B_s^0$ decays to common final states and is related to the decay width difference 
$\Delta \Gamma^s$  between the two mass eigenstates by,
\begin{eqnarray}
\Delta \Gamma^s & = & \Gamma_L^s - \Gamma_H^s =  2\,  |\Gamma_{12}^s| \cos  \phi_s \ ,
\end{eqnarray}
whereas the mass difference in Eq.~(\ref{qp}) is proportional to the off-diagonal element $|M_{12}^s|$,
\begin{eqnarray}
  \Delta M^s & = &  M_H^s - M_L^s =  2\,|M_{12}^s| \ ,
\end{eqnarray}
and equals the frequency of the $B_s^0 - \bar B_s^0$ oscillations. 

The two mass eigenstates of $B$ mesons are expected to be almost pure $CP$ eigenstates when $\Delta \Gamma/\Gamma$ is negligible, which is the 
case for $|\Gamma_{12}^s| \simeq 0$ and $\Gamma^s_{12}/M_{12}^s$ is approximately real in the Standard Model (since $\phi_s$ is very small).
By virtue of Eq.~(\ref{qp}) it follows that $q/p \simeq - M_{12}^{s*}/|M_{12}^s| =  \exp (i\phi_M)$ and $|q/p|= 1$ (to within 1\%~\cite{Nakamura2010}).  However, 
experiment seems to indicate that $\Delta \Gamma^s/\Gamma^s$ could be as large as 22\% for the $B_s^0$~\cite{Abazov:2011ry}. We shall return to this 
point shortly in Section~\ref{sec2}.

The time-dependent $CP$-violating asymmetry for $B_s^0 \to J/\psi \phi$ is defined to be,
\begin{eqnarray}
  a_{J/\psi \phi}(t)  & = & \frac{\Gamma (\bar B_s^0(t) \to J/\psi \phi) - \Gamma (B_s^0(t) \to J/\psi \phi)}{\Gamma (\bar B_s^0(t) \to J/\psi \phi) +\Gamma (B_s^0(t) \to J/\psi \phi)} \ ,
\end{eqnarray}
and can be shown to be written as~\cite{Anikeev:2001rk}, 
\begin{eqnarray}
 a_{J/\psi \phi}(t)  & = & - \left [(1 - |\lambda_{J/\psi \phi}|^2) \cos (\Delta M^s t) + 2\, \mathrm{Im}\, \lambda_{J/\psi \phi} \sin (\Delta M^st)\right ] \! / (1 + | \lambda_{J/\psi \phi} \ |^2),
\end{eqnarray}
provided $\Delta \Gamma^s/\Gamma^s$ is small and $|q/p|=1$. We made use of the definition,
\begin{eqnarray}
  \lambda_{J/\psi \phi} & = & \eta_{J/\psi \phi}\ \frac{q}{p} \,\frac{\bar \mathcal{A}(\bar B_s^0\to J/\psi \phi) }{\mathcal{A}(B_s^0\to J/\psi \phi) }  \ ,
\end{eqnarray}
where $\mathcal{A}$ denotes the complex decay amplitude, $\bar \mathcal{A}$ is the $CP$ conjugate amplitude and $\eta_{J/\psi \phi}=\pm 1$ describes $CP$-even and 
-odd components in the final state,  the separation of which requires an angular analysis~\cite{Dighe:1995pd}. For the so-called third type of $CP$ violation with and without 
mixing, one has in addition to $|q/p| =1$ also the condition $|\bar \mathcal{A}/\mathcal{A}| = 1$,\footnote{\, In the decay $B_s^0 \to J/\psi \phi$, the decay amplitude 
$\mathcal{A}$ receives penguin contributions from the two CKM terms $V^*_{ub} V_{us} \propto \lambda^ 4$ and $V^*_{cb} V_{cs}\propto \lambda^2$ with 
$\lambda = 0.2257$ in the Wolfenstein parametrization~\cite{Nakamura2010}. The tree diagram also contains the weak phase $V^*_{cb} V_{cs}$ so that the penguin 
amplitude proportional to $V^*_{ub} V_{us}$ is CKM suppressed and additionally subleading in the strong coupling $\alpha_s$. As a consequence, neglecting this term, 
the dominant tree and  penguin amplitudes effectively contribute a single overall weak phase which leads to $|\bar \mathcal{A}/\mathcal{A}| = 1$ and no {\em direct\/} 
$CP$ violation occurs.} so that $|  \lambda_{J/\psi \phi} | =1$:
\begin{eqnarray}
   a_{J/\psi \phi}(t)  & = & -  \mathrm{Im}\, \lambda_{J/\psi \phi}\,  \sin (\Delta M^st) = \eta_{J/\psi \phi} \, \sin (2\beta_s) \sin (\Delta M^st) \ .
\end{eqnarray}
Hence, the asymmetry directly measures the phase differences between particular CKM matrix elements and introduces no uncertainty due to strong interaction phases,
which are often of non-perturbative origin and not well known; namely, the strong interaction effects all cancel exactly since $|\lambda_{J/\psi \phi}| =1$ or at least
very close to 1. Since $\beta_s$ is predicted to be very small in the Standard Model, no appreciable $CP$ violation should be detected in experiment. Therefore,
any large deviation from this prediction could indicate new physics (NP) contributions manifest in the modified mixing phase:
\begin{eqnarray}
  2\,\beta_s & = &  2\,\beta_s^\mathrm{SM} - \phi_s^\mathrm{NP} .
\end{eqnarray}

\section{The related decay $\bm{B_s^0 \to J/\psi\, f_0(980)}$ \label{sec2}}

First experimental determinations of $\beta_s$ have come from the CDF~\cite{Aaltonen:2007gf} and D$\emptyset$~\cite{Abazov:2008fj} Collaborations and 
their initial values hinted at a possible large deviation (of order $2.2\sigma$ if both results are combined) from the Standard Model. The latest D$\emptyset$
measurements~\cite{Abazov:2011ry} of $\Delta \Gamma^s$ and $\beta_s$ are only marginally smaller than those in~\cite{Abazov:2008fj} and in particular 
$\Delta \Gamma^s = 0.163^{+0.065}_{-0.064}~$ps$^{-1}$.  However, an updated CDF measurement seems to be more consistent~\cite{Giurgiu:2010is} with 
the  Standard Model values. One is hopeful that LHCb will provide tighter constraints on $\beta_s$ and $\Delta \Gamma^s$ in the near future.

It was already mentioned that $CP$ violation can be measured using angular analyses since the final state $J/\psi\, \phi$ is not a $CP$ eigenstate. 
This requires  more events to acquire a similar sensitivity to that obtained if the decay proceeds solely via $CP$-even or $CP$-odd channels. The related 
channel $B_s^0 \to J/\psi\, f_0(980)$ is in that sense advantageous, as the decay is to a single $CP$-odd eigenstate and does not require an angular analysis.
As in the case of $B_s^0 \to J/\psi\phi$, its $CP$ violating phase is given by $-2\beta_s$ up to higher corrections. However, as for other scalar mesons,
the exact flavor and constituent content is not known to date.\footnote{\, Notwithstanding popular descriptions of scalar mesons as molecular bound-states of 
mesons as well as tetra-quarks, a Dyson-Schwinger equation approach to the scalar $\bar qq$ ground state with a non-perturbative 
kernel beyond the rainbow-ladder truncation determines the mass of the flavor-pure scalar to be $m_{\sigma} = 900$~MeV~\cite{Chang:2011ei}. A more 
sophisticated description in terms of a mixing angle between $\bar uu(\bar dd)$ and $\bar ss$ components has not been realized yet but is feasible. 
Further improvements include pion- and kaon-loop effects in the non-perturbative kernel which can shift the pole mass by about 8\%~\cite{Holl:2005st}. 
In short, viewing scalar mesons, such as the $f_0(980)$, exclusively as a $\bar qq$ or  ${\bar q}^2 q^2$ state may simply be too naive~\cite{Pennington:2007eg}.}
The mass of the $f_0(980)$ is well estimated at  $m_{f_0} = 980\pm 10$~MeV, yet its width is only poorly known due to the nearby opening of the $\bar KK$ 
channel and its dependence on a given final state. It can be rather narrow in the case of $B^{\pm} \to K^{\pm} f_0(980)$ decays and estimates of the width 
are in the range $40-100$~MeV~\cite{Nakamura2010}. For a general overview of scalar mesons we refer to the review by C.~Amsler  {\em et al\/}.
in the Particle Data Group book~\cite{Nakamura2010} and references therein.

It has also been argued that the angular analysis in the decay $B_s^0 \to J/\psi\phi$  is complicated by the $J/\psi\, f_0(980)$ channel, since it contributes $S$-wave 
$K^+K^-$ pairs which can interfere with those originating from the $\phi$. This $S$-wave should also be manifest in the appearance of $f_0(980) \to \pi^+\pi^-$
decays. Following this observation, the level of $S$-wave ``contamination''  was  proposed to be estimated by the following ratio~\cite{Stone:2008ak}:
\begin{eqnarray}
\label{Rf0phi}
\mathcal{R}_{f_0/\phi} & = &\frac{\Gamma (B_s^0 \to  J/\psi f_0(980),  f_0(980) \to \pi^+ \pi^-)}{\Gamma(B_s^0 \to  J/\psi \phi, \phi \to K^+ K^-)}\ .
\end{eqnarray}
Initial estimates based on similar decays in the charm sector show this ratio to be of the order of $20\%-30\%$~\cite{Stone:2008ak}. These estimates rely on 
experimental data on $D_s^+ \to f_0(980) \pi^+$ and $D_s^+ \to \phi \pi^+$ decay rates and seem to indicate that the $S$-wave contribution of $f_0(980)\to K^+K^-$ 
cannot be ignored when analyzing the angle $\beta_s$ in $B_s^0 \to J/\psi \phi$. Likewise, Xie {\em et al\/}. found the effect of an $S$-wave component on 
$-2\beta_s$ to be of the order of $10\%$ in the $\phi$ resonance region~\cite{Xie:2009fs}.  A first calculation of this ratio based on decay amplitudes derived
in QCD factorization (QCDF) and a model calculation of the non-perturbative transition amplitude  $\mathcal{A}(B_s \to f_0(980))$~\cite{Leitner:2010fq,ElBennich:2008xy}  
are discussed in the following sections.

A treatment of possible $S$-wave contributions in experimental analyses was presented in a recent analysis of $\beta_s$~\cite{Giurgiu:2010is}. 
In there, the CDF Collaboration finds that $S$-wave contribution within $\pm 10$ MeV about the $\phi$ meson is less that 6.7\% at 95\% confidence level.
Incidentally, these preliminary CDF results~\cite{Giurgiu:2010is} on $\beta_s$ and $\Gamma^s$ point at a reconciliation with the Standard Model
values and it remains to be clarified which impact on the analyses the additional $S$-wave contribution have.

\section{Nonperturbative aspects in decay amplitudes}

The details of the $B_s^0 \to  J/\psi f_0(980)$ and $B_s^0 \to  J/\psi \phi$ decay amplitudes calculated in QCDF can be found in 
Ref.~\cite{Leitner:2010fq}. We here concentrate on the nonperturbative matrix element that emerges from the factorized amplitude. The amplitude of interest 
is the heavy-to-light transition between a $B_s^0$ and the scalar $f_0(980)$ meson which decomposed into Lorentz invariants gives rise to two form factors,
\begin{equation}
\label{defbs}
\langle f_0(p_2)| \bar s \, \gamma_{\mu}(1-\gamma_5) b | \bar  B_s^0(p_1)  \rangle =
  \Bigl( p_\mu  - \frac{m_{B_s^0}^2-m_{f_0}^2}{q^{2}}q_\mu \Bigr ) F_1^{B_s^0\to f_0}(q^{2}) 
   +   \frac{m_{B_s^0}^2 -m_{f_0}^{2}}{q^{2}} q_{\mu}\ F_{0}^{B_s^0\to f_0}(q^{2}) \ ,
\end{equation}
with $p_1^2 = m_{B_s^0}^2$, $p_2^2= m_{f_0}^2$, $q= p_1 - p_2$ and $p=p_1+p_2$. The matrix element of the $B_s^0 \to \phi$ transition amplitude is
given by a similar decomposition and introduces five more form factors~\cite{Leitner:2010fq}.

Common relativistic quark-model approaches represent heavy-to-light transition amplitudes by triangle diagrams, a 3-point function between the Bethe-Salpeter 
amplitudes (BSA) of a heavy $(H)$ and a light $(M)$ meson and the weak  coupling represented by the transition amplitude $\langle M(p_2) | \bar q\, \Gamma_I h | H(p_1) \rangle$.  
This is the  generalized impulse approximation, which in the language of Dyson-Schwinger equations~\cite{nucl-th/0703094} is the leading term in their systematic and 
symmetry preserving truncation:
\begin{eqnarray}
  \mathcal{A}(p_1,p_2) = \mathrm{tr} \int\! \frac{d^4k}{(2\pi)^4} \, \bar \Gamma_{M}^{(\mu)}(k;-p_2)  S_q(k+p_2) \Gamma_I(p_1,p_2)  S_Q (k+p_1)  \Gamma_H(k; p_1) S_{q'} (k) \ ,  
 \label{heavylightamp}
\end{eqnarray}
where $S(k)$ are (dressed) quark propagators, $Q=c,b$; $q=q'=u,d,s$;  $\Gamma_M$  is the light meson BSA with  $M=S,P,V,A$ and the index $\mu$ indicates its possible vector 
structure. $\Gamma_I= \gamma_\mu (1-\gamma_5)$ or  $\sigma_{\mu\nu} q^\nu(1+\gamma_5)$ is the interaction vertex and $\Gamma_H$ is the heavy meson BSA. The trace is 
over Dirac and color indices. Calculation of these matrix elements in lattice-regularized QCD or with QCD sum rules follows a somewhat different approach based 
on heavy-light correlation functions. A brief review on effective and non-perturbative approaches to heavy-light transition form factors and in particular discrepancies 
between model predictions at large time-like momentum transfer, $q^2$, can be found in Ref.~\cite{ElBennich:2009vx} 

As  discussed earlier, the lack of a precise constituent picture of the scalar $f_0(980)$ makes a precision calculation of the transition matrix element in Eq.~(\ref{defbs})
impossible for the time being. Nevertheless, the $B_s \to f_0(980)$ form factors have recently been obtained in QCD sum rules~\cite{Ghahramany:2009zz,Colangelo:2010bg} 
and pQCD~\cite{Li:2008tk}  for $q^2=0$, where an extrapolation to the value $F_{0,1}^{B_s^0 \to f_0}(m_{J/\psi}^2)$ is required. They have also been calculated 
by some of us~\cite{ElBennich:2008xy} from the constituent quark three-point function, the vertices of which are the weak interaction coupling, 
$\gamma_\mu(1-\gamma_5)$, and two BSA for the $B_s$ and $f_0(980)$ mesons. In this relativistic dispersion-relation model, a phenomenological parametrization 
of the $B_s$ is obtained from the simultaneous calculation of the weak decay constant $f_{B_s}$ (known from lattice-QCD simulations). In an attempt  to find a 
suitable form of the $f_0(980)$ BSA, we constrained the mixing angle between strange and non-strange $\bar qq$ components and appropriate width parameters 
by means of experimental data on the branching fractions of the decays $D_{(s)}\to f_0 (980) P$ with $P=\pi, K$. The scalar $F_0^{B_s^0 \to f_0}(q^2)$ and vector 
$F_1^{B_s^0 \to f_0}(q^2)$ form factors are then obtained for any physical time-like momentum transfer $q^2$ and no extrapolation is needed. We recall that only 
the vector form  factor $F_1^{B_s^0 \to f_0}(q^2)$ enters the decay amplitude $\mathcal{A} (B_s^0 \to  J/\psi f_0(980))$. We deduce from the extrapolation parametrization in 
Ref.~\cite{Colangelo:2010bg} that $F_1^{B_s^0 \to f_0}(m_{J/\psi}^2)\simeq 0.3$, which is compatible with our prediction $F_1^{B_s^0 \to f_0}(m_{J/\psi}^2) \simeq 0.4$
\cite{ElBennich:2008xy} within theoretical errors.

\section{The ratio $\bm{\mathcal{R}_{f_0/\phi}}$ and new physics contributions in the light of recent measurements}

\begin{figure}[t]
\begin{center}
\includegraphics*[scale=0.6]{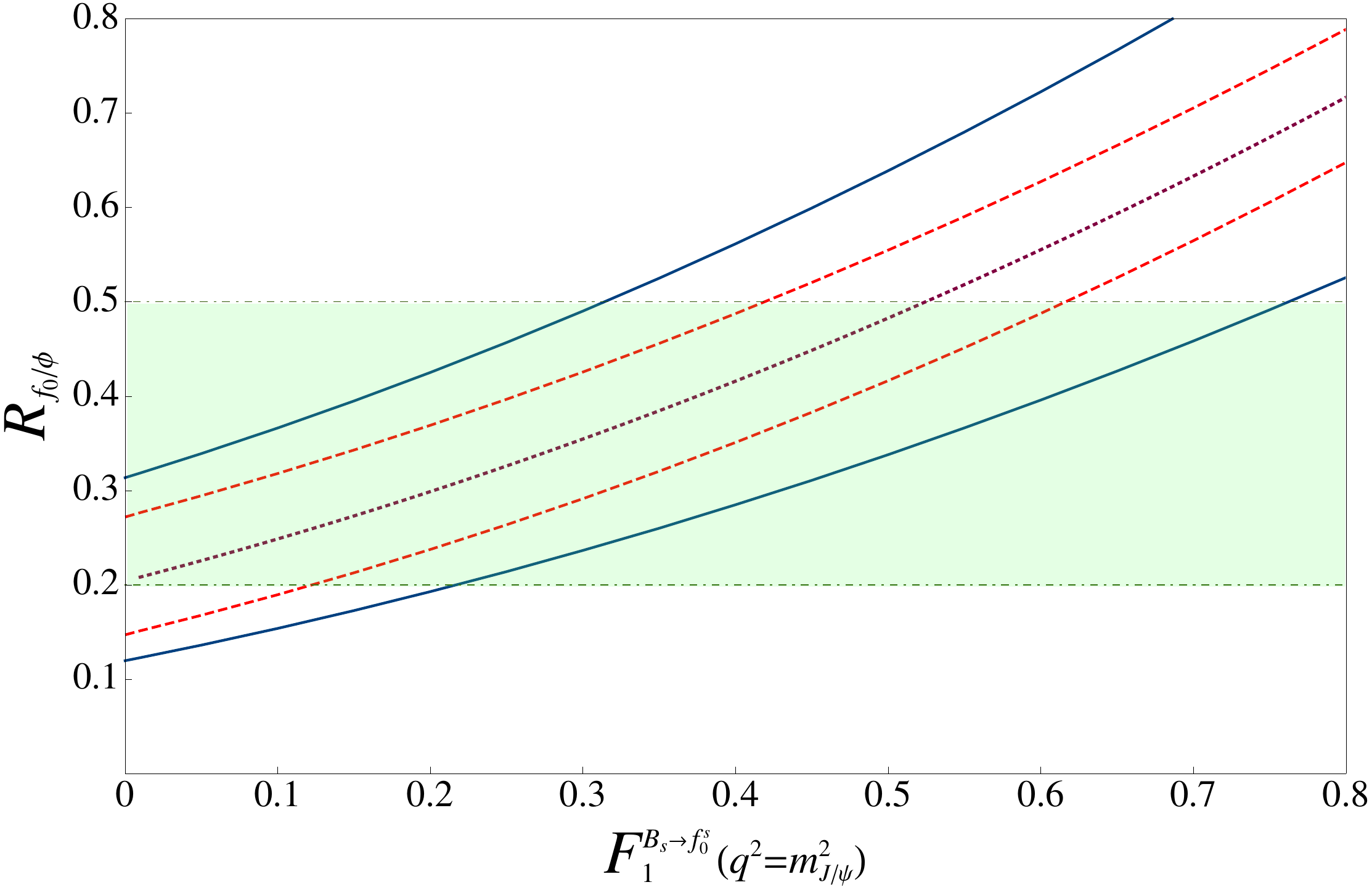}
\caption{The ratio $\mathcal{R}_{f_0/\phi}$ as a function of the transition form factor $F_1^{B_s^0 \to f_0^s}(m_{J/\psi}^2)$ based on the $B_s^0 \to  J/\psi f_0(980)$ 
and $B_s^0 \to  J/\psi \phi$ decay amplitudes in Eqs.~(2) and (4) of Ref.~\cite{Leitner:2010fq} for $\zeta^{(h)}=0$. The area between the two dashed lines accounts 
for the uncertainty of the decay constants ($f_{B_s} =260 \pm 30$ MeV and $\bar f_{f_0}=380 \pm 40$ MeV) while the solid lines include in addition the uncertainties 
on the decay rates $f_0(980) \to \pi^+ \pi^-$ and $\phi \to K^+ K^-$. The single dotted line is the prediction for the central values of the decay constants. The shaded 
area between the two dot-dashed horizontal lines represents the window of experimentally motivated estimates for $\mathcal{R}_{f_0/\phi}$
\cite{Stone:2008ak,Ecklund:2009fia}. \label{fig1} }
\end{center}
\end{figure}

\begin{figure}[t]
\begin{center}
\includegraphics*[scale=0.6]{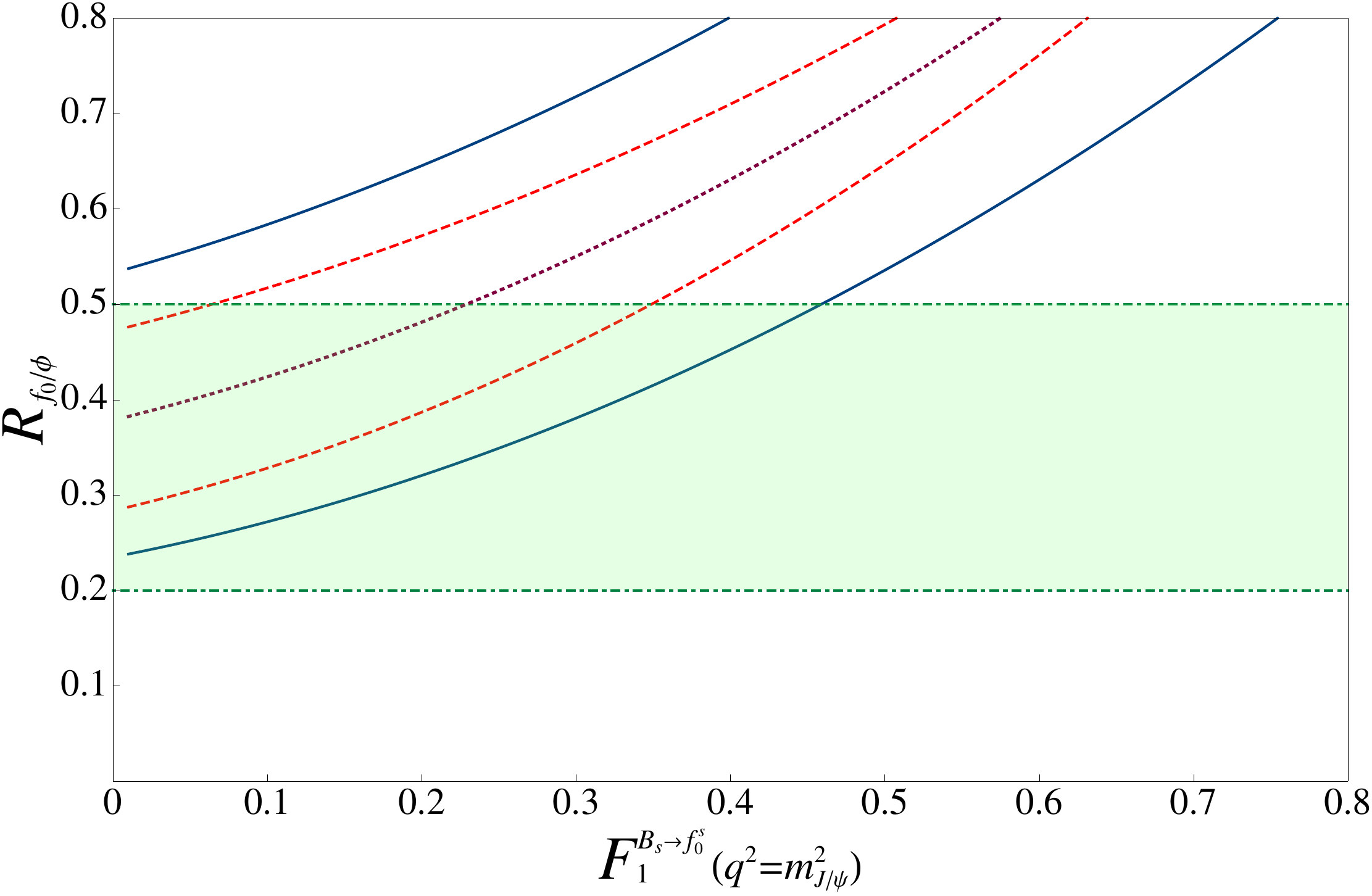}
\caption{The ratio $\mathcal{R}_{f_0/\phi}$ as in Figure~\ref{fig1} but including possible new physics contributions $\zeta^{(h)}$ in Eqs.~(2) and (4) of 
 Ref.~\cite{Leitner:2010fq}.  \label{fig2} }
\end{center}
\end{figure}

Equipped with an estimate of the $F_1^{B_s^0 \to f_0}(m_{J/\psi}^2)$ form factor value and the QCDF expression for $\mathcal{R}_{f_0/\phi}$ in Eq.~(\ref{Rf0phi}),
we can predict this ratio in dependence of different inputs in the  $B_s^0 \to  J/\psi f_0(980)$ and $B_s^0 \to  J/\psi \phi$ decay amplitudes. To account for 
the main source of uncertainty, we do so by plotting this ratio as a function of  $F_1^{B_s^0 \to f_0}(m_{J/\psi}^2)$ in Figs.~\ref{fig1} and \ref{fig2}. 
Two other sources of uncertainties are included in both figures which are due to the theoretical errors of the vector decay constant, $f_{B_s}$, the scalar 
decay constant, $\bar f_{f_0}$,\footnote{\, See Section~IV in Ref.~\cite{Leitner:2010fq} for a discussion of the scalar decay constant $\bar f_{f_0}$ defined by  
$m_{f_0} \bar f_{f_0} = \langle 0 | \bar q q | f_0  \rangle$ and its relation to the vector decay constant $f_{f_0}$.} and the experimental error on the decay 
rates  $f_0(980) \to \pi^+ \pi^-$~\cite{ElBennich:2008xy, Ecklund:2009fia}  and $\phi \to K^+ K^-$~\cite{Nakamura2010}.
We plot the evolution of $\mathcal{R}_{f_0/\phi}$ for a rather large window of form factor values, though the domain of interest due to the theoretical estimates
\cite{ElBennich:2008xy,Colangelo:2010bg} is in the range $F_1^{B_s^0 \to f_0}(m_{J/\psi}^2) \simeq 0.3 - 0.4$. As seen in Figure~\ref{fig1}, in this particular range 
the central value of $\mathcal{R}_{f_0/\phi}$ is about $0.35\pm 0.05$ which is compatible with the estimates based on the ratio of (differential) decay rates, 
$\Gamma (D_s^+ \to f_0(980) \pi^+ )/\Gamma (D_s^+ \to \phi \pi^+)$~\cite{Stone:2008ak,Ecklund:2009fia}, depicted by the shaded area.

Additional short-distance amplitudes $\zeta^{(h)}$,  where $h$ denotes the helicity of the $J/\psi\phi$ state, proportional to the dominant CKM term were 
introduced in the QCDF amplitudes for $B_s^0 \to  J/\psi f_0(980)$ and $B_s^0 \to  J/\psi \phi$~\cite{Leitner:2010fq}. In a sense, these phenomenological 
amplitudes mock up other or beyond Standard Model physics contributions to the decays. In addition, we assume that whatever these other amplitudes are, 
they contribute equally to all helicities states of $J/\psi\, \phi$ and to $J/\psi f_0(980)$, as they originate in the same flavor-changing neutral  current, $b\to s \bar cc$, 
of the corresponding penguin diagrams: $\zeta^{(h)}_{J/\psi \phi} = \zeta_{J/\psi \phi} = \zeta_{J/\psi f_0}$. Generically, we write the amplitudes as,
\begin{equation}
    \mathcal{A} =  | \mathcal{A}^{\mathrm{SM}} |  \, e^{2i \beta_s^{\mathrm{SM}}}    +  | \mathcal{A}^{\mathrm{NP}} | \, e^{i(2 \beta_s^{\mathrm{SM}} - \phi_s^{\mathrm{NP}})}  
                          =   | \mathcal{A}^{\mathrm{SM}} |  \, e^{2i \beta_s^{\mathrm{SM}}} \left ( 1+ \mathcal{R}  \, e^{-i \phi_s^{\mathrm{NP}}}  \right ) \ ,
\end{equation}
with $\mathcal{R} =  | \mathcal{A}^{\mathrm{NP}} /  \mathcal{A}^{\mathrm{SM}} |$. The amplitudes $\mathcal{A}^{\mathrm{NP}} \sim \zeta^{(h)}$ are adjusted so they 
give the best possible agreement with experimental data on  $B_s^0 \to J/\psi \phi$, which includes the branching ratio, the longitudinal, parallel and perpendicular 
polarization fractions, $f_L, f_{\parallel}$ and $f_{\perp}$,  and two relative phases, $\phi_{\parallel}$ and $\phi_{\perp}$~\cite{Nakamura2010,Abazov:2008jz,Kuhr:2007dt}. 
They are then inserted in the $B_s^0 \to  J/\psi f_0(980)$ decay amplitude. The evolution of $\mathcal{R}_{f_0/\phi}$ including the $\zeta^{(h)}$ amplitudes as 
a function of $F_1^{B_s^0 \to f_0}$ is presented in Figure~\ref{fig2} from which it obvious that the ratio is strongly enhanced. For the same range as previously, 
$F_1^{B_s^0 \to f_0}(m_{J/\psi}^2) \simeq 0.3 - 0.4$, the central value of $\mathcal{R}_{f_0/\phi}$ is of the order $0.6\pm 0.05$. Taking into account the theoretical 
uncertainties this is still within an acceptable range from the estimates in the shaded area.

On the other hand, recent first measurements of the ratio $\mathcal{R}_{f_0/\phi}$, consistent with each other, seem to favor our calculation for $\zeta^{(h)} =0$ 
and a form factor $F_1^{B_s^0 \to f_0}(m_{J/\psi}^2) < 0.4$:
\begin{eqnarray}
\mathcal{R}_{f_0/\phi} & =  & 0.275 \pm 0.041 \pm 0.061 \quad (\mathrm{D\emptyset~Collaboration}~\cite{Abazov:2011hv})  \; , \nonumber \\
\mathcal{R}_{f_0/\phi} & =  & 0.257 \pm 0.020  \pm 0.014 \quad (\mathrm{CDF~Collaboration}~\cite{Aaltonen:2011nk})  \; , \label{rf0phiexp} \\
\mathcal{R}_{f_0/\phi} & =  & 0.252^{+0.046+0.027}_{-0.032-0.033} \quad  (\mathrm{LHCb~Collaboration}~\cite{Aaij:2011fx})   \; ,  \nonumber
\end{eqnarray}
where in each case the errors are statistical and systematic, respectively. The calculated value $F_1^{B_s^0 \to f_0} \simeq 0.4$~\cite{ElBennich:2008xy}
actually leads to a bigger ratio, $\mathcal{R}_{f_0/\phi} = 0.42$, which rises to $0.63$ when the $\zeta^{(h)}$ amplitudes are included. Clearly, as Figure~1 
instructs us, if the decays $B_s^0 \to  J/\psi f_0(980)$ and $B_s^0 \to  J/\psi \phi$ are merely due to Standard Model interactions (save for $\Lambda_{\mathrm{QCD}}/m_b$
corrections neglected in our decay amplitudes based on QCDF), then $0.15-0.2$ is a more likely value for $F_1^{B_s^0 \to f_0}$. If, however, these
decays receive contributions from a yet unknown source then it is not plausible that they should be equal in magnitude and phase for both final states
unless one admits a rather unrealistic value for $F_1^{B_s^0 \to f_0} < 0.1$. Even if this was the case, the new physics amplitudes $\mathcal{A}^{\mathrm{NP}}$
must necessarily contribute different phases to the $B_s^0 \to  J/\psi f_0(980)$ and $B_s^0 \to J/\psi \phi$ amplitudes which results in interferences 
so that  $\mathcal{R}_{f_0/\phi}$ comes close the observed values in Eq.~(\ref{rf0phiexp}).

\section{Acknowledgment}

B.~El-Bennich would like to thank the organizers, Amand F\"assler and Jochen Wambach, of the Erice International School of Nuclear Physics ``From Quarks 
and Gluons to Hadrons and Nuclei''  for their invitation and the pleasant atmosphere during the school. We appreciated valuable communication with Sheldon Stone 
and Craig Roberts.  This work was supported by FAPESP grants nos.~2009/53351-0, 2009/51296-1 and 2010/05772-3 and CNPq grant no.~306395/2009-6.

\end{document}